\documentclass[11pt]{article}
\usepackage{a4,epsfig}
\newcommand{\tifor}[2]{{

\medskip \noindent \rule{18cm}{0.4mm}}

{\vspace{1ex} \noindent}

{\noindent \Huge\bf#1}

{\vspace{1ex} \noindent}  

{\noindent \Large\it#2 $^{*}$}

{\vspace{2ex} \noindent \rule{18cm}{0.4mm}
\par \vspace{3ex}}}

\newcommand{\refust}{ \begin{itemize}}

\newcommand{\refusl}{\end{itemize}}

\newcommand{\refst}{ \begin{enumerate}}
\newcommand{\reffo}[5]{{\footnotesize\sf \item #1
{\footnotesize\sl #2} #3 {\bf #4} #5}}
\newcommand{\refsl}{\end{enumerate}}

\newcommand{\rammen}[2]{\begin{tabular}[t]{|p{81mm}|}
\hline
\multicolumn{1}{|l|}{} \\
\multicolumn{1}{|c|}{#1} \\
\multicolumn{1}{|l|}{} \\
#2 \\
\hline
\end{tabular}}

\begin{document}
\raggedbottom
\sloppy
\hbadness=5000
\twocolumn[\tifor 
{The  LHC, shining light  on the  Dark Side}
{Anna ~Lipniacka, University of Bergen, Norway}]

\par
\noindent
{\bf 

Starting in the summer of 2007, 
the  Large Hadron Collider (LHC) will collide
proton beams at  center-of-mass energies  of
14 TeV  exceeding by a factor of ten what was previously achieved.
It will be located in the 27km long underground tunnel, in which
the Large Electron Positron collider (LEP) was working until the year 2000.
The Large Hadron Collider  is  a part of the accelerator complex 
of the European Laboratory
of Particle Physics (CERN)$^{(1)}$, situated  on the  
Franco-Swiss border close to Geneva.}

Those who read the Economist$^{(2)}$
do not need to be convinced that physics is the queen of experimental
sciences.  Physics pertains to the whole 100\% of the content of the Universe,
while only 5\% of it is ordinary matter, and can be the subject of Chemistry.
The so-called Cold Dark Matter forms close to 25\% of the content, the rest is
the even more mysterious Dark Energy.

Dark Matter consists of heavy, elusive particles, interacting with ordinary
matter even less than neutrinos do. However it interacts gravitationally, and
was most probably responsible for the formation of Large-Scale Structures.
VIRGOHI21 is a recently discovered$^{(3)}$ Dark Matter galaxy, where ordinary
matter is only 0.1\%.  In most of the galaxies including ours the density of
Dark Matter is about ten times larger than the density of ordinary matter.
The stars in our Galaxy would fly apart, if Dark Matter was not gluing them
together with gravitational forces. The local density of it is close to 0.3
proton masses/${\rm cm}^3$. A cup of coffee contains around five Dark Matter
particles.  However, the coffee cup contains at the same time around $2
\times 10^{28}$ protons and neutrons (provided it is filled with coffee)!
This illustrates the problem of studying Dark Matter particles here on
Earth. Yet, as the Big Bang produced apparently so much Dark Matter, perhaps
we can produce lots of it too, if we recreate similar conditions.

\vspace{1ex}
\noindent
\rule{8.6cm}{0.1mm}
\vspace{1ex}
\noindent
{\footnotesize \sf $^{*}$Fysisk institutt, email:anna.lipniacka@ift.uib.no 
}

Recreating energy densities needed for Dark Matter production will be one of
the purposes of proton-proton collisions at the Large Hadron Collider.  The
rest energy of a proton is 0.938 GeV.  To attain such an energy an {\it
electron} has to be accelerated in a potential difference of 938 million volts!
A proton in the LHC beam will have the energy of 7000 GeV, thus 7500 times
more than its rest energy. Should LHC protons race to the Moon against a beam
of light, they would arrive only 2.7 meters behind. Beams will be organized in
bunches of around $10^{11}$ protons each, colliding 40 million times per
second in several collision points. The energy stored in the beam is close to
$10^{8}$ Joules. If you have to dump all of it in an instant you will
evaporate an equivalent of 300 kg of water!

Three detectors are being built to register proton-proton collisions, two
general purpose detectors ATLAS (A Toroidal LHC ApparatuS)$^{(4)}$ and CMS (Compact
Muon Spectrometer), and a more specialized one, LHCb, dedicated to physics
involving beauty quarks.  The fourth detector, ALICE, will study heavy ion
collisions.  Beams of heavy nuclei (ions) will occupy around 15\% percent of
the beam-time starting from the third year of the LHC running.  Groups in
Norway are active in the ATLAS and ALICE Collaborations.  The physics programs
of ATLAS and ALICE are quite different. This article is devoted to physics
topics pertaining to proton-proton collision.
 
The construction of detectors is progressing fast. Figure~1
shows the ATLAS detector filling up its grand underground cavern.  Parts of the
detector are  already in place and
being tested with muons originating from cosmic rays.


\begin{figure}[htb]
\begin{center}
\begin{minipage}[t]{0.9\linewidth}
\mbox{\epsfig{file=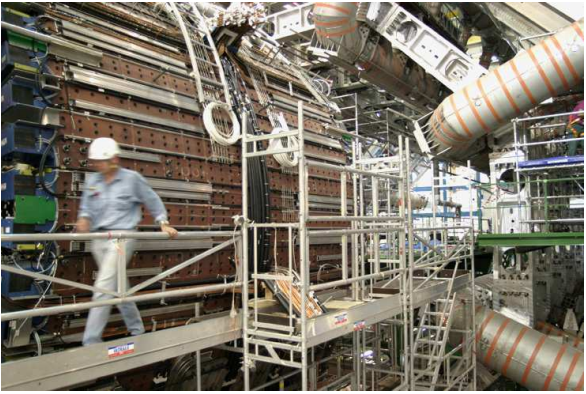,width=\linewidth}}
\end{minipage}
\end{center}
\caption{
The ATLAS cavern in October 2005, insertion of
the calorimetric systems.
The coils of the ATLAS toroidal
magnetic system are also visible.}
\label{fig:figure1}
\vfill
\end{figure}

To probe small structures and to produce new particles, high energies in
elementary collisions are needed.  Protons are not elementary however, and
collisions between the constituents of protons: quark and gluons, will
occur. These constituents carry on average considerably less energy than the
proton, rendering genuine high energetic collisions less probable. That is why
lots of collisions are needed.  Several proton-proton collisions will occur
every 25 nanoseconds.  Extremely short time spacing between subsequent beam
interactions poses extreme requirements on the detector technology, local data
transmission and storage. Every 25ns a stream of particles, produced in beam
collision will flash the detector, moving with nearly the speed of light to
the outer layers.  Before particles from one collision reach the outer layers
there will be another flash of particles coming from the next collision.
On-detector electronics will have to correctly associate signals from
particles to the right collision event. The pixel detector, the most granular
subdetector of ATLAS, provides 80 million bits of information which needs to
be correctly handled!  The data transmission rates involved are gigantic. The
ATLAS detector will transmit more data than all of the worlds phone networks
integrated.

Only one collision in a million will be interesting enough to deserve
permanent storage and further study.  The selection system, the so-called
trigger, will have to reliably reject one million of ``spam'' collisions to
find the interesting single one to store, all these in an extremely small
fraction of a second. Imagine a spam-mail filter performing a similar task!
Two hundred megabytes of interesting data will be stored by the ATLAS
experiment 100 times per second. This corresponds to 1200 CDs per minute.  To
reconstruct and analyze this information GRID-based technology linking PCs and
PC farms all over the world will be used. CERN-related research in Norway in
particular is a forefront runner in GRID technology development.

The search for collisions where Dark Matter
is produced (see Fig.~2)  
is just one of the topics the ATLAS experiment will embark on.
One of the enigmas ATLAS and CMS  will be trying to solve   
is why the ``ordinary particles'', the known elementary
fermions and bosons have masses at all.


\begin{figure}[htb]
\begin{center}
\begin{minipage}[t]{0.8\linewidth}
\mbox{\epsfig{file=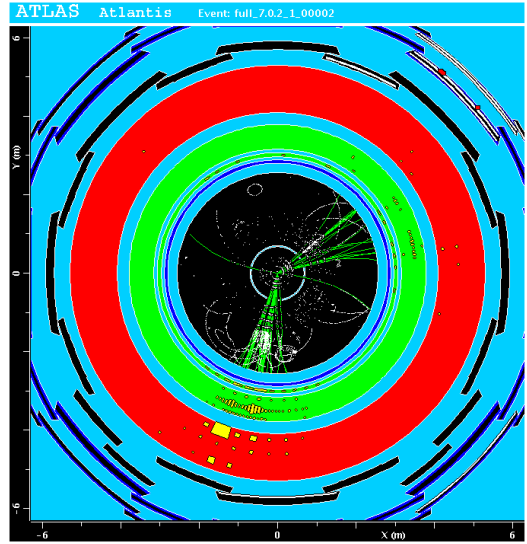,width=\linewidth}}
\end{minipage}
\end{center}
\caption{ \footnotesize{
Simulated ATLAS event with the production of a Dark Matter candidate particle.
The Dark Matter particle does not interact with the ATLAS detector and
can be ``seen'' as a recoil against streams of ordinary particles
registered in the detector. 
}}
\label{fig:figure2}
\vfill
\end{figure}

{\it The most viable hypothesis of mass generation is 
that particles acquire masses
via interaction with the so-called Higgs field, which permeates the vacuum.
The particles sort of glue themselves to the Higgs field.
Thus, the heavier the particle, the stronger is the interaction with
the Higgs field. 
The presence of the field in the vacuum causes the so-called Electroweak
Symmetry breaking: a difference between electromagnetic and weak
interactions, and between, for example, an electron and a neutrino.}
There is a vast literature on the subjects touched in this short article, 
which can be tackled starting for example from this reference ${^(5)}$. 

The Higgs mechanism was proposed by Peter Higgs of Edinburgh University (see
Figure~3).  Before going into details of explanations, let me suggest how the
Higgs field hypothesis can be verified.  Einstein postulated in 1905 that
electromagnetic field should have its quantum, the photon ({$\gamma$}). It is
now believed that all interactions and fields should manifest themselves as
quanta $\rightarrow$ particles.  ATLAS will hunt for the quantum of the Higgs
field, the Higgs boson.  The Higgs boson must interact strongly with heavy
particles, thus it is expected to decay mostly to them. 


\begin{figure}[htb]
\vspace{-3mm}
\begin{center}
\begin{minipage}[t]{0.60\linewidth}
\mbox{\epsfig{file=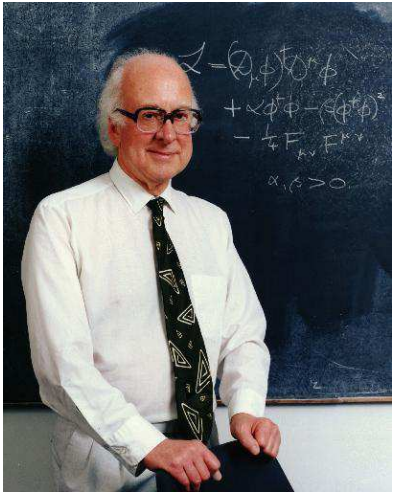,width=\linewidth}}
\end{minipage}
\end{center}
\caption{\footnotesize{Peter Higgs was born in Bristol in 1929, 
and is presently professor emeritus at
the University of Edinburgh.
In the  mid-1960s Higgs and others proposed a mechanism to give  masses 
to elementary particles, which explains the Electroweak Symmetry breaking
at the same time. The  Higgs field, and  the quantum of the Higgs field, 
the Higgs boson
are inherent to this explanation. 
Peter Higgs' lecture, ``My life as a boson'', 
is available
from most on-line encyclopaedias.
}}
\label{fig:figure3}
\vfill
\end{figure}

The Higgs particle is the only missing piece of the so-called Standard Model.
This quite unfortunate name arose before particle physicists realized they
were going to be stuck with the Standard Model for decades! The Standard Model
describes ``ordinary'' matter particles, fermions, and interactions between
them transmitted by bosons.  It has been shown to work in an extremely
accurate way up to center-of-mass energies of the order of 100 GeV, explaining
an era of the order of a nanosecond after the Big Bang.  It allowed to make
very accurate calculations, which were confirmed experimentally with matching
precisions.  However, there is no candidate for the Dark Matter particle in
the Standard Model, there is really no good place for the neutrino masses, and
even the missing piece of the puzzle, the Higgs boson, does not fit in by
itself without invoking beyond-the-Standard Model particles.  Thus physics
``beyond the Standard Model'' must exist.\\

\noindent
{\Large \bf The Standard Model}\\

Known matter is built of fermions, while interactions keeping it together
are transmitted  by bosons.

Fermions can be ``Lego blocks'' of interesting structures because the Pauli
principle forbids that all of them fall to the lowest energy state. Electrons,
protons and neutrons are needed to build atoms.  Protons and neutrons contain
other fermions: up and down quarks. Electrons, the electron neutrino, and up
and down quark form the so-called first fermion family  needed
to build ordinary matter.  Surprisingly, two more families exist. The second
family consists of the muon neutrino, the muon, charm and strange quarks,
while the third one has the tau neutrino, the tau, top and bottom quarks.

When it comes to all known  properties (except masses) the second and
third families seem to be replicas of the first. The masses of fermions are
quite well measured, again except for neutrinos. The masses of the first
family fermions are a small fraction of the proton mass. The top quark mass,
however, is equal to that of 187 protons, it is the heaviest known elementary
particle. It is also, unlike the proton, point-like and has no internal
structure down to around 1/100th of the proton size.  Thus masses of family
members increase with the family number, although this might be different for
neutrinos. The reason for mass patterns and mass values of elementary fermions
remains a mystery.
\medskip

Increase a hydrogen atom to the size of a  physicist's
office in Norway, and 
the proton will become the  size of a  small dust particle. If the 
proton moves, the  change of  its position is communicated to the electron
via electromagnetic interactions. This information propagates with the speed
of light, as an electromagnetic wave.

The static electric field surrounding all charged particles becomes
electromagnetic waves when the particle accelerates.  Einstein's concept of
electromagnetic wave particle, the photon, was a revolutionary idea. Nowadays
electromagnetic waves can be observed as photons by anybody equipped
with a photo-diode or a photomultiplier.  Electromagnetic interactions are
mediated by photons, electrically neutral bosons with spin=$\hbar$ and zero
rest mass.  Other known elementary bosons are eight electrically neutral
massless gluons transmitting strong interactions, heavy electrically charged
$W^+,W^- $ and the electrically neutral $Z^0$ transmitting the Weak
Interactions.  The $Z^0$ weighs as much as about 97 protons.

The experimental observation of the W and Z bosons brought the
Nobel Prize to Carlo Rubbia and Simon Van der Meer of CERN, in 1984.  In the
years 1989-1995 the LEP accelerator at CERN produced a few million Z
bosons, allowing for precise studies of the Weak Interactions.\\

\newpage
\noindent
{\Large \bf The Electroweak Symmetry and its breaking}\\

All electrically-charged particles 
feel electromagnetic interactions. All known 
fermions have Weak Interactions charge. Actually, from the point of view of
Weak Interaction there is no difference between an electron and a neutrino.
In the so-called natural units $(\hbar=c=1)$ the  electron (proton) charge is
a magic number $e=-0.303$ (0.303), 
while the weak charge of every 
elementary fermion
is  close to another magic number of 
$g=0.631$. 
If both the electron and the  neutrino
were massless, and  if all bosons were massless 
as well there would be hardly
any observable difference between the electron and 
the neutrino and between Weak and Electromagnetic
interaction. Weak attraction would be as strong as  
the electromagnetic one, and neutrinos would be captured into bound 
states in atoms as much 
as electrons are. This is the world of ``unbroken Electroweak Symmetry'' 
in particle physicists' jargon.   
However, in our world the Electroweak Symmetry
is broken: the neutrino and the electron are perceived as different particles, 
and Weak Interaction
bosons are massive, while the photon is massless. As a result 
Weak Interactions  are 
much weaker than the
electromagnetic ones at  atomic scales.

It is tempting to uncover the Electroweak Symmetry and find a common
description of electromagnetic and weak $\rightarrow$ Electroweak
Interactions. Such a unified Electroweak theory exists due to Sheldon
Glashow, Steven Weinberg and Abdus Salam who were awarded the Nobel Prize in
1979.  The Electroweak symmetry is extremely useful and actually allows to
perform precise calculations in the Electroweak theory.  How to deal with the
apparent breaking of the symmetry while preserving all its good features? The
answer is the Higgs mechanism.  A heuristic analogy with a ferromagnet is
often used to explain it, see Box 1.

\begin{figure}[htb]
\begin{center}
\begin{minipage}[t]{0.60\linewidth}
\mbox{\epsfig{file=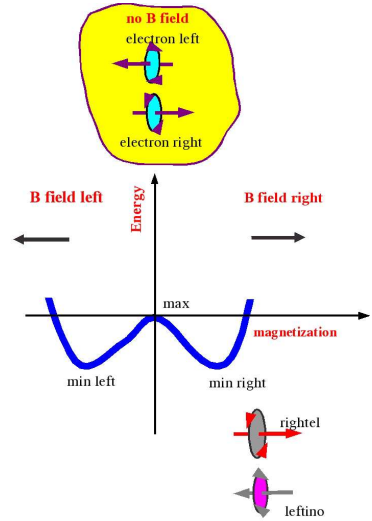,width=\linewidth}}
\end{minipage}
\end{center}
\vspace{-5mm}
\caption{ {\footnotesize
An electron can have two magnetic moment states, one parallel, one
anti-parallel to a chosen direction in space.
The state of a
ferromagnet without the magnetic field is symmetric, there is no chosen
direction in space, electrons with magnetic moment
pointing to the left and right  have the same energy. 
However the lowest energy
state of the ferromagnet is with a non-zero value of the magnetization.
If minimum on the right is chosen
resulting  spontaneous magnetic field would 
point to the right. {\it This is the so-called spontaneous symmetry breaking}.
The result is that {\it rightels} (electrons
with magnetic moment to the right) and {\it leftinos} (electrons with magnetic moment to the left)
behave like  different particles and  there
is a non-zero ``expectation'' value of magnetic field in the ferromagnet.
}}
\label{fig:figure5}
\vfill
\end{figure}

\noindent
\rammen{\bf Box 1. The Higgs mechanism}{
\vspace{-0.5cm}
\footnotesize
\noindent

The analogy makes use of concepts of magnetic fields, electron magnetic moment
and ferromagnets.  Electrons are charged and have spin ($ {\frac{1} {2}}
\hbar$).  
An electron can have two magnetic moment states, one parallel, one
anti-parallel to a chosen direction in space (see Figure~4).
Most electric phenomena are not affected by the
direction of the electron's magnetic moment. The electron mass does not depend
on it.  Imagine, that the space around us is
permeated by a constant magnetic field pointing from left to right on 
figure~5. In this case all observed electrons would have their
magnetic moment pointing to the left, assuming minimal energy in the
magnetic field.  Let's call them {\it leftinos}.  Electrons with magnetic
moment to the right ({\it rightels}) would have to be
produced by providing some energy to leftinos.  In other words some energy
would be needed to flip the magnetic moment of an electron in the magnetic
field, but most probably the ``true'' picture of the situation would be hard
to uncover. Rightels could then decay into a photon and a leftino.
Rightels and leftinos would appear as different particles with different
masses and different interactions.
}

\newpage
\noindent
\rammen{\bf Box 1 cont.}{
\vspace{-0.5cm}
\footnotesize

The reason behind the magnetic field in the vacuum could be that that it has
properties of a ferromagnet, and its lowest energy state is with a non-zero
magnetic field pointing in some direction, as in Figure~4.  
The state of a
ferromagnet without the magnetic field is symmetric, there is no chosen
direction in space, rightels and leftinos have the same mass and interactions,
and are simply the same particle, the electron.  However the lowest energy
state of the vacuum is with a non-zero value of the magnetic field pointing in
some direction.  {\it This is the so-called spontaneous symmetry breaking}.
The result of it is that rightels and leftinos are different particles, there
is a non-zero ``expectation'' value of magnetic field in the vacuum.  This is
an approximate picture of the Higgs mechanism, if one translates magnetic field
into Higgs field, rightels and leftinos to electrons and neutrinos.

\hspace{1mm}
There is one more analogy here.  It is enough to heat  the
ferromagnet to allow it to go to the higher energy state and destroy 
the spontaneous
magnetization. The symmetry is recovered. We expect  this was the situation a
fraction of a nanosecond after the Big Bang.  When the Universe was cooling
down, the vacuum chose its lowest energy state and the Electroweak symmetry
broke. In a ferromagnet there can be domains, where the symmetry is broken in a
different way and the magnetization points to different directions. Do we have
a similar situation in the Universe? Possibly, but this
is a subject for another article.

\hspace{1mm}
There is lots of proofs of the Electroweak theory with spontaneous symmetry
breaking. The most striking one is that it relates the ratio of the weak to
the electromagnetic charge to the ratio of the masses of the W and Z
bosons. This relation was confirmed experimentally. When (if?)  the Higgs
particle is found the picture of Electroweak symmetry breaking will be
complete.

}

\vspace{5mm}

A lot is known about the Higgs boson, even if it was not found yet. Actually
its mass is the only unknown parameter.  Even the mass has been ``measured''
already with a certain accuracy.  The Higgs boson has to be heavier than
114~${\rm GeV}/c^2$ otherwise it would have been observed at LEP.  From its
``shadow'' in the masses of other particles: the top quark and the W boson,
one can infer it should be lighter than about 200~${\rm GeV}/c^2$.  How?
Atomic physicists are familiar with the Lamb shift. Willis Eugene Lamb
received the Nobel Prize in 1955 for his discoveries concerning the fine
structure in hydrogen. The Lamb shift, hyperfine energy splitting between S
and P orbitals in hydrogen is caused by the creation of virtual
electron-positron pairs in the atomic electric field (vacuum polarization) and
modification of both the electron mass and the magnetic moment due to
interaction with quanta of the atomic electric field (virtual photons).  In a
similar way the masses of the top quark and the W boson are affected by the
existence of the Higgs boson.  This method of determining a particle mass
without producing it was already tested in the 90's. The LEP experiments
determined the top quark mass with an accuracy of 10\%, before it was actually
observed at the Tevatron experiments in the Fermi National Laboratory near
Chicago, USA.\\

\noindent
{\Large \bf Beyond the Standard Model}\\

Symmetry is one of the basic concepts of science. Nature and art are full of
symmetries.  Many of us pondered on the symmetric beauty of snowflakes and
wondered about the laws of nature.

Symmetries are often only approximate. Left-right symmetry is an example.
External features of our bodies are to a large extent left-right symmetric.
Internally however, having the liver on the right hand side and the heart on
the left we strongly violate the left-right symmetry. The situation is quite
similar in the world of elementary particles. Weak interactions strongly
violate the left-right (P) symmetry and matter-antimatter exchange symmetry
(C), while all other interactions seem to conserve them. Lewis Caroll's Alice,
walking through a looking glass into a room where left and right were reversed
might have found herself in a completely different Universe, in which certain
nuclear interaction do not occur.  Our Universe is believed to be
CPT-symmetric. Exchange at the same instant left with right (P), matter with
antimatter (C), and reverse the arrow of time (T), and nothing observable will
change, even if each of these operations performed separately produces a
Universe different from ours.

Certain symmetries are woven into the structure of space and time. We can
change summer to winter time without rewriting physics books. We can also move
the zero longitude from Greenwich to Bergen, and formulations of all know
physics laws will remain the same.  Emmy Noether (see Fig.~5)
proved in 1915 that every continuous symmetry of physics (time and
space translation and space rotations are examples) results in a conserved
physical quantity. For example, space translation symmetry results in the
conservation of momentum. Noether's theorem linking symmetries to conservation
laws is one of the basic foundations of physics.  Thus, it might be
instructive to examine even approximate symmetries.


\begin{figure}[htb]
\begin{center}
\begin{minipage}[t]{0.60\linewidth}
\mbox{\epsfig{file=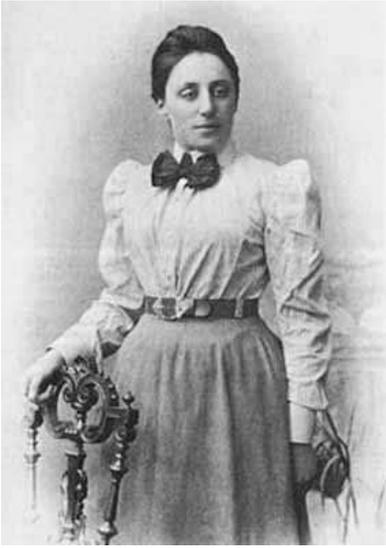,width=\linewidth}}
\end{minipage}
\end{center}
\caption{ \footnotesize{Emmy Noether 1882-1935, born in Erlangen, Germany. 
Noether's theorem stating that every continuous symmetry in physics results in
a conserved physical quantity is a cornerstone of physics. 
Her most famous work ``Die Invariante Variationsprobleme''
appeared in 1918.  
Noether
was granted a doctorate in 1907 at the
University of G{\"o}ttingen, and later lectured in Vienna and in Italy. In 1915
Hilbert and Klein invited her to return to G{\"o}ttingen.  Hilbert was
advertising her courses under his name, as she was not allowed to give courses
under her own.  
Much of her work appears in
papers written by her colleagues and students, 
rather than under her own name.}}
\label{fig:figure4}
\vfill
\end{figure}

The so-called supersymmetry (SUSY) might be one of them. SUSY is a
boson-fermion symmetry.  The known matter particles are fermions, while
interaction particles are bosons. If the world was supersymmetric each fermion
would have an identical mirror particle, but a boson, and vice versa. The
names for these mirror particles are already there, the bosonic partners of
fermions are called sfermions, while fermionic partners of the photon, W, Z,
and gluon were named photino, gluino, Wino, Zino. Is there anything more than
names? We have not found any supersymmetric particles so far. If sfermions
were identical to fermions except for spin, their masses would be the same as
well, and we would have seen them already. Thus if supersymmetry is real, it
must be, like the left-right symmetry, only approximate. SUSY partners must be
heavier than ``ordinary'' fermions and bosons.  Why do we need them at all?

As symmetries go, aesthetics is one of the arguments.  SUSY is consistent with
the Theory of Relativity and Quantum Mechanics.  It also helps to integrate
gravity with other interactions.  The only known candidates for a quantum
theory of gravity, string theories, are supersymmetric, they produce equal
numbers of bosonic and fermionic particles, however at a very high energy
scale.

There are also more ``practical'' arguments in favor of SUSY.  If
supersymmetric partners are lighter than about 1000~${\rm GeV}/c^2$, Weak,
Electromagnetic and Strong forces become equally strong at distances of around
$10^{-32}$ m.  The already mentioned ``Lamb effect'' (vacuum
polarization), is responsible for the apparent change of electric, weak and
strong charges of particles with the distance. Depending on the distance from
a particle we see more or less polarized vacuum on our way toward it. How the
vacuum is polarized and what effect this has on the observed charge depends on
the strength of the field and on what particle-antiparticle pairs exist in the
real world!  SUSY brings in the right particles, and all charges become equal
if viewed from a distance of $10^{-32}$m! Like in the Lamb effect, the
existence of virtual particles in the vacuum affects not only the observed
charges but also the masses. One can imagine that every particle drags behind
itself a cloud of virtual particles it interacts with. This has disastrous
effects for the Higgs boson mass. It glues itself so strongly to virtual top
quarks, that its mass becomes much larger than experimentally preferred
bounds.  If SUSY exists, the supersymmetric partner of the top quark, the
stop, would have a healing effect on the Higgs boson mass, by partially
screening the boson from interactions with virtual top quarks.  Another nice
feature of SUSY became clear some time after it was conceived. The lightest
supersymmetric particle is just an ideal Dark Matter candidate.

If SUSY is in the energy range needed to provide all the nice features
above, it will be pretty quickly  discovered by the ATLAS and CMS experiments. 
Our understanding of the content of the Universe
will improve from 5\% to 30\%!
\medskip

Observable space around us has three dimensions. One speaks often about time
as the fourth dimension, with different properties. If there are more
space-like dimensions, they must be of a small size, otherwise we would have
observed them by now. String theories invoke extra space dimensions of the
Planck-size, about $10^{-35}$m, far too small to be observed experimentally in
the foreseeable future. Two small extra dimensions would be like a small
sphere was attached in every point of our space.  Recently it was noted that
extra spatial dimensions of sub-millimeter size could solve a long standing
problem why gravity is so much weaker than other interactions. If we allow
gravity (and no other interactions or particles) to propagate into these extra
dimension it simply ``leaks out'' of our space, and we do not see its full
strength.  However, if the energies high enough to probe distances comparable
with the size of the extra dimensions are attained, the full strength of
gravity will be revealed.  We might see its consequences in the form of Black
Hole nanotechnology: production of nanosized Black Holes at the LHC.\\

\noindent
{\Large \bf Concluding remarks}\\

What is the use of supersymmetric particles, extra space dimensions, Black
Holes and Dark Matter?  Before the electron was discovered any questions about
its possible ``utility'' might have been equally difficult to
answer. TV, electricity and the World Wide Web are all by-products of basic
research.  Perhaps old science-fiction authors' dream of storing energy in
Black Holes and tunneling via extra dimensions to other parts of the Universe
will become true?  One thing is sure. The quest for universal answers is an
inherent part of human culture and the technological development is often a
by-product.  The LHC is pushing the technology frontier in areas of
electronics, computing, telecommunication, detectors and accelerators,
superconducting magnetic systems and cryogenics.  Last, but perhaps not
least, after the LHC experience particle physicists will be able to construct
the best spam-mail filters in the world!\\

\noindent
{\Large \bf Acknowledgements}\\

 A version of this paper is due to appear in ``Fra fysikkens Verden'', 
a journal
of Norwegian Physical Society. The author would like to
 thank Per Osland for useful comments and for
his efforts to remove all signs of Polish from the text. \\

\noindent
{\Large \bf Bibliography}
\refst

\itemsep -0.1cm

\reffo{www.cern.ch}{}{}{}{}

\reffo {The Economist:}{362 (8254)}
{``With All Thy Getting, Get Understanding p. 12 `` ,''Survey: The Universe, 
Dark for dark business'', pp. 47-58 }{2002}{}

\reffo {R.F. Minchin et al:}
{ ``A Dark Galaxy in the Virgo Cluster Imaged at 21-cm'',}
{astro-ph/0508153 v1, Mon. Not. R. Astron. Soc. 23, September 2005}{}{(2005)}

\reffo{atlas.ch}

\reffo {C. Quigg:} { Gauge Theories of the Strong, Weak, and Electromagnetic
Interactions. }
{Addison-Wesley, 1997}{}{}

\refsl


\end{document}